\documentclass[a4paper,twoside]{article}

\usepackage{epsfig}
\usepackage{subfigure}
\usepackage{calc}
\usepackage{amssymb}
\usepackage{amstext} 
\usepackage{amsthm}
\usepackage[table,xcdraw]{xcolor}
\usepackage[fleqn]{amsmath} 
\usepackage{pslatex}
\usepackage{enumitem}
\usepackage{textcomp}
\usepackage{natbib}
\usepackage{url}
\usepackage{array,multirow,graphicx}

\usepackage{SCITEPRESS}      

\begin{document}

\title{
Towards Automated Management and Analysis of Heterogeneous Data Within Cannabinoids Domain
}


\author{ 
\authorname{Kenji Koga\sup{1}, Maria Spichkova\sup{2} , Nitin Mantri\sup{3} 
}
\affiliation{ 
\sup{1} iSelect, Cheltenham, Australia, kenji.koga@iselect.com.au
} 
\affiliation{ 
\sup{2}School of Science, RMIT University, Melbourne, Australia, maria.spichkova@rmit.edu.au
} 
\affiliation{ 
\sup{3}School of Science, RMIT University, Melbourne, Australia, nitin.mantri@rmit.edu.au
} 
}

\abstract{Cannabinoid research requires the cooperation of experts from various field biochemistry and chemistry to psychological and social sciences. The data that have to be managed and analysed are highly heterogeneous, especially because they are provided by a very diverse range of sources.
A number of approaches focused on data collection and the corresponding analysis, restricting the scope to a sub-domain. Our goal is to elaborate a solution that would allow for automated management and analysis of heterogeneous data within the complete cannabinoids domain. The corresponding integration of diverse data sources would increase the quality and preciseness of the analysis.
In this paper, we introduce the core ideas of the proposed framework as well as present the implemented prototype of a cannabinoids data platform. \footnote{Preprint. Accepted to the 14th International Conference on Evaluation of Novel Approaches to Software
		Engineering (ENASE 2019). Final version published by SCITEPRESS, http://www.scitepress.org}
}

\onecolumn 
\maketitle 
\normalsize \vfill

%===============================================
\section{\uppercase{Introduction}}
\label{sec:introduction}
 
Since the beginning of the first phytocannabinoids characterisation in the 20\textsuperscript{th} century, see \cite{grotenhermen2004clinical}, and the first studies using tetrahydrocannabinol and cannabidiol, we faced a boost in research involving cannabinoids. With the use of medicinal cannabis being legalised in a growing number of countries, several studies in different health areas have been conducted such as in inflammatory diseases, see e.g., \cite{Hasenoehrl2017}, neurological disorders or related symptoms, see \cite{solimini2017neurological,Pertwee3353}, cancer, see \cite{Pagano2017,naderi2018effects,pastor2004therapy,milano2017recent} and cardiovascular diseases, see  \cite{mendizabal2007cannabinoids}. 

To conduct the cannabinoid research effectively and efficiently, data from different sources have to be considered. For example, a researcher interested in finding the best treatment for a particular disease has to analyse data on specific cannabinoid strains, the treatment data and the corresponding effects that patients or doctors have described. 
Since these data are handled by different individuals and institutions, which generally have their own data format, this task requires the integration of multiple data sources that are heterogeneous. Moreover, the diversity of the user backgrounds requires the corresponding adjustments of the system interface. 

A number of approaches aimed to combine data in areas such as pharmacology, see \cite{gray2014,drugbank2017,Samwald2011}, and health sciences, see \cite{Puppala2015,Reis2018}. 
Most of them applied some open standards like OpenEHR\footnote{\url{http://openehr.org/}} to present the collected data.
This solution is not applicable for the case of cannabinoids research: 
we are dealing not with a single domain that has already their data standards, but we have to collect and integrate
 data from multiple heterogeneous sub-domains. Thus, the challenges are not only in the integration of data collected from a single sub-domain, e.g. health data records, but also in integration of multiple heterogeneous domains.

\emph{Contributions:}
In this work we propose a platform for automated collection, management and analysis of cannabinoids data. This platform will integrate data from several cannabinoids data sub-domains in order to provide means for higher quality research analysis. 
We also present the implemented prototype of the proposed platform. 

\emph{Outline:} The rest of this paper is organised as follows. 
Background and related work are discussed in Section \ref{sec:related_work}. %  
Sections \ref{sec:architecture} and \ref{sec:impl} introduce the core ideas of the proposed cannabinoid data platform as well as its current implementation. 
Section \ref{sec:conclusions} concludes the paper and provides some future work directions.

% 
%===============================================
\section{\uppercase{Background and Related Work}}
\label{sec:related_work}

In the cannabinoids domain, there are % 
a number of projects focusing on
the acquisition of cannabinoid data such as Strainprint\footnote{\url{http://strainprint.ca}}, SeedFinder\footnote{\url{https://en.seedfinder.eu}}
and Open Cannabis Project\footnote{\url{http://opencannabisproject.org}}. 
The Strainprint project collects personal data (user profile), data on strains, ingestion methods and dosage. Its core objective is to keep track of the effectiveness of treatments.  
The other two projects focus on collecting and sharing information about cannabis strains, 
without any objective to integrate these data with any other type of data  to identify the most effective treatment using cannabinoids.

\cite{Sawler2015} analysed the strain data classification using DNA samples which showed that strain names do not represent genetically unique variety. Samples with identical strain names were more genetically similar to samples with different names than to identical ones. This demonstrates another issue that has to be covered when developing a system for  management and analysis of  cannabinoids data: 
researchers cannot rely on strain names only, the genetic similarity has to be taken into account.

In genomics, Pharmacogenomics Knowledgebase, see \cite{whirl2012pharmacogenomics}, and Public Health Genomics Knowledge Base, see \cite{yu2016knowledge}, are open Web-based knowledge bases that collect, curate and provide information about human genetics and population health. They focus on providing high-quality information to support medicine-implementation projects and population health, respectively. To achieve this, they periodically extract data from, e.g. scientific publications, using manual, natural language processing and Machine Learning techniques. They differ from our approach because they are not fully automated as well as they are focused on a single domain.

In what follows we discuss the approaches that do not focus on cannabinoids domain, but present some computer science concepts related to our research -- big data analytics in healthcare, cloud-based solutions for health information systems, etc. 
\cite{luo2016big} conducted a literature review on 
big data application in biomedical research and health care, focusing on the big data application in four major biomedical sub-disciplines:   bioinformatics,  clinical informatics,  imaging informatics, and   public health informatics. The authors identified 68 relevant papers, and their study demonstrated 
\emph{``While big data holds significant promise for improving
health care, there are several common challenges facing all the
four fields in using big data technology; the most significant
problem is the integration of various databases.''} To provide an effective and efficient solution for this problem is one of the goals of the platform we propose.

%Wang et al. 
\cite{Wang2015,Wang2018} presents a survey on 26 big data analytics cases in healthcare research field and derived some of the best practices. This analysis resulted in an architecture with 5 logical layers including data collection, data aggregation, analytics, information exploration and data governance. In the Data collection layer, they have all data sources collection such as structured, semi-structured and unstructured data. Data aggregation layer deals with data extraction and transformation. In the Analytics layer, they process and analyze data using, for example, MapReduce and data mining. 

MapReduce is a programming paradigm and an associated implementation for processing and generating large datasets, see \cite{dean2008mapreduce}. MapReduce can be also seen as the core of   Apache Hadoop\footnote{\url{https://hadoop.apache.org}}, an open source platform for the distributed processing of structured, semi- and unstructured data.  
In Information exploration layer, \cite{Wang2015,Wang2018} proposed to generate reports, alerts and notifications outputs derived from the Analytics layer. In Data governance layer, they  propose to deal with ethical, legal, and regulatory challenges managing all the life-cycle of data, security, privacy and policies. In our work, we derived data heterogeneity architectural layers following these best practices adapted to the cannabinoid data domain.

A model of a cloud-based platform and its open-source implementation was presented and refined in \cite{yusuf2015chiminey},\cite{yusuf2017chiminey}, \cite{spichkova2016towards},  \cite{spichkova2015scalable} and \cite{spichkova2016managing}.
This model allows researchers to conduct experiments  requiring complex computations over big data. This platform might be integrated within Analytics and Visualisation Layer of the architecture we propose, see Section~\ref{sec:architecture}.

\cite{calabrese2015cloud} reviewed
main cloud-based healthcare and biomedicine applications, especially  focusing on healthcare, biomedicine and bioinformatics solutions. The authors summarised core issues and problems related to the use of such platforms for the storage and analysis of patients data.

\cite{Bahga2015} developed and extended Cloud Health Information Systems Technology Architecture, which allows clinical data integration, access and analytics. It achieves integration through mapping source-specific format to a domain model. It supports formats, such as Health Level-7 messages and Clinical Document Architecture and raw ASCII. Once data schema matches the domain model, the framework proceeds with a parallel MapReduce aggregation and transformation task and write the result to HDFS storage. Data access and analytics can be performed through Hadoop ecosystem components such as Pig\footnote{\url{https://pig.apache.org}} or  Hive\footnote{\url{https://hive.apache.org}} providing seamless access to all the data inside the cloud using HCatalog from Hadoop. Since we will deal with different kinds of domain standards, we will have to research which are the main standards available in the cannabinoid data research area and provide equivalence of semantics. However, we will take advantage of a similar cloud infrastructure approach in order to get all the benefits that it provides such as parallelization of processing jobs, fault-tolerance and scalability.
 
eClims, see \cite{Savonnet2016}, is another example of an integration framework to deal with data and schema variability in Biomedical Information Systems (BIS). Since Biomedical research area has to deal with the constant integration of increasing number of databases and ontologies, they have created eClims to facilitate the integration of new data and extend data models at the same time assuring quality using Databases and Semantic Web theory. In this approach, the authors preferred a manual solution for semantic analysis of collected data that were collected from several providers, hence had a different structure. The question of a full automation was still open in that approach.

There are also several approaches to integrate heterogeneous pharmacology data such as the Life Sciences Linked Open Data (LSLOD) cloud, but it stills a difficult task to acquire relevant results. It is necessary to combine knowledge generated from drugs, physiological functions in biological systems and underlying biological interactions. This demands efforts on integrating multiple heterogeneous sources, perform manual entity reconciliation and disambiguation, which are non-trivial and non-scalable tasks. \cite{Kamdar2017} developed a Platform for Linked Graph Analytics in Pharmacology to perform integration of four different data sources from LSLOD cloud, using a data model, to abstract all the relevant mechanisms of drug relations, query federation, where SPARQL\footnote{\url{https://www.w3.org/TR/rdf-sparql-query}} queries are performed in all sources to generate k-partite network and probabilistic model to discover associations between, for example, drugs and adverse drug reactions. In our work, we will take advantage of their infrastructure on how to deal with Semantic Web Technologies, i.e., their data model and query federation in order to perform data integration.

\begin{figure*}[ht!]
\begin{center}
\includegraphics[width=1.0\textwidth]{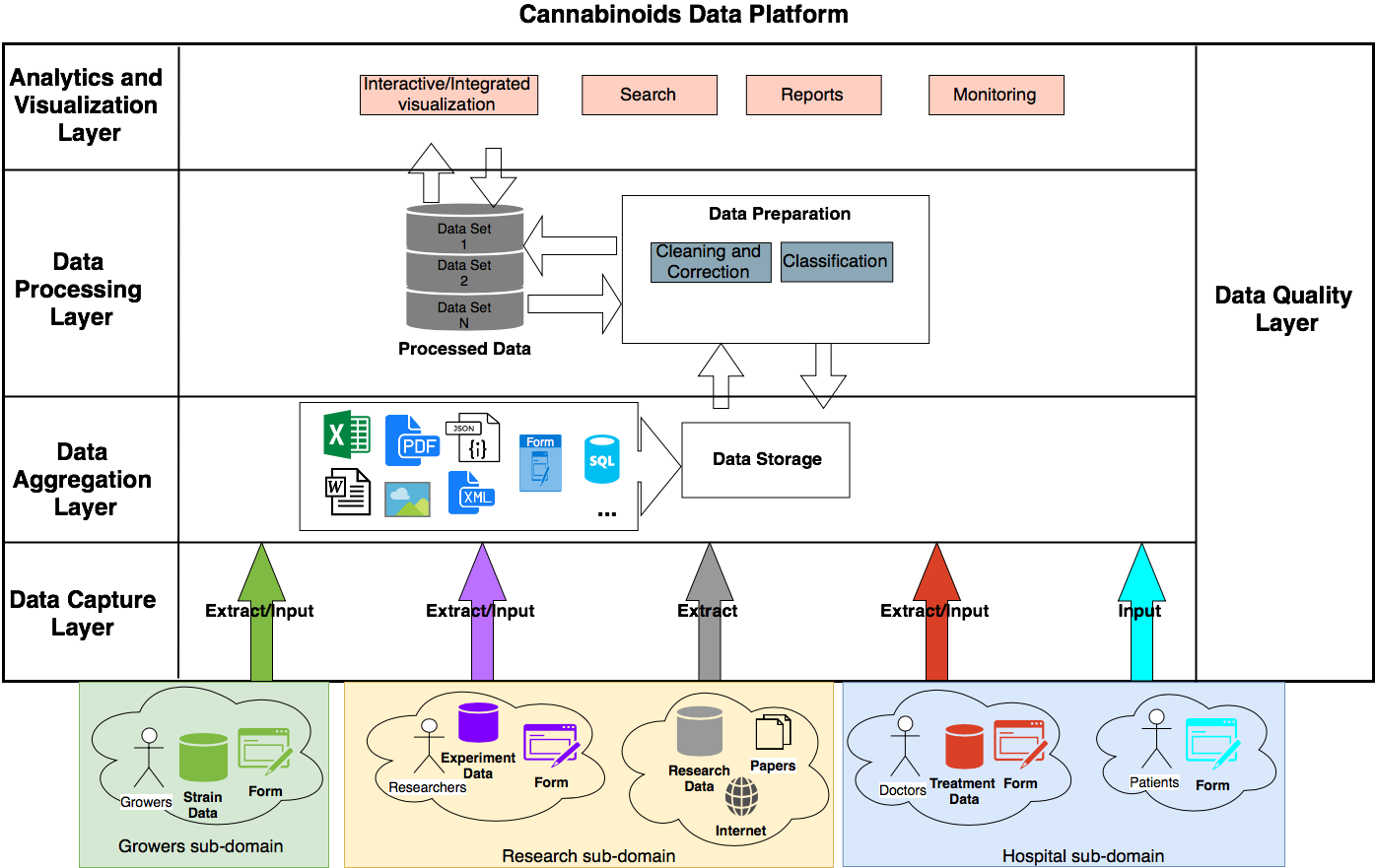}
\\~
\caption{Architecture of the CDP}
\label{fig:arch}
\end{center}
\end{figure*}

% 
%===============================================
\section{\uppercase{Proposed Architecture}}
\label{sec:architecture}

In this section, we discuss the core aspects of our Cannabinoids Data Platform (CDP) for the management and analysis of cannabinoids data. Figure \ref{fig:arch} presents a layered architecture of the CDP. 

In the \emph{Analytics and Visualization Layer}, all the concepts regarding user interaction are handled. Users should be able to interact with the CDP through different kinds of interfaces. We analyze and provide corresponding functionalities required for each user type. Interactive and integrated visualizations are provide in different options for visualizations in charts and tables. 
To implement an efficient search functionality, data has to be prepared and indexed (in \emph{Data Processing Layer}). Like in the \emph{Data Capture Layer}, we have to consider here the specifics of Cannabinoids Domain to provide an easy-to-use interface that allows  to find, order/rank, match and analyse the collected data. 

As the functionalities provided within this layer are mostly focused on researchers, it makes sense to apply the technologies that are already used in the corresponding research community. Thus, to support the analysis and visualisation of the large amount of collected data, we are going to apply a research-oriented cloud computing platform Chiminey that provides user-friendly interface for the computation/ analysis set up, as well as visualisation of the calculation results as 2D or 3D graphs, see \cite{yusuf2015chiminey} and \cite{spichkova2015scalable}. 
Monitoring and alerts has to be provided to notify users about updates in the data they are interested (through a previous subscription), for example, new treatment data or changes in experiments.  

In the \emph{Data Processing Layer}, data are prepared and provided in the repository of processed data. 
This repository contains a number of data sets, each of them is specialised in providing data for a specific usage. 
This is achieved through prior data cleaning and correction, where useless data is removed, i.e., data that is not useful for a specific data set context, and corrected if there is a possibility to do so. 
Other important steps are data classification, categorization and indexing, as they provide a basis for better search results as the data with similar features are kept together in an appropriate set and structured using Semantic Web technologies such as presented in \cite{Kamdar2017}. 
The processing infrastructure used here is similar to the one proposed by \cite{Bahga2015}. 
In the \emph{Data Aggregation Layer}, data of any different format in the cannabinoids domain are aggregated and stored. We are considering storage of some of the unstructured (Excel, Word, PDF and images), semi-structured (JSON and XML) and structured (MySQL and PostgreSQL) data since we do not have full knowledge of all available data types in the cannabinoid domain.

In the \emph{Data Capture Layer}, data is extracted from research data bases as well as publications or manually fed in by the users (growers, researchers, doctors/ treatment coordinators, and patients). The users can provide their data in several ways, for example  uploading files or typing information in a Web form (in the case of researchers, doctors / treatment coordinators, and growers) or using a mobile application (in the case of patients). 
The user interface for patients should be as straightforward and simple as possible, as some of the patients might have not much experience in using mobile applications.  

Since each user of the platform can provide their data using different mechanisms and interfaces, some of the data might be unstructured, depending on the sub-domain:
\begin{itemize}
    \item 
    The data collected within the hospital and growers sub-domains is always structured, as all the users within these sub-domains can provide the data only using the corresponding forms / interfaces. 
    Thus, if we would limit the platform to these sub-domains, a data warehouse solution woulds be sufficient, see \cite{george2015data}. 
    \item 
    The vast part of the data collected within the research sub-domain is unstructured and data extraction becomes a challenge within this sub-domain. This means that only the data lake solution is applicable, see e.g., \cite{soini2017efficient,miloslavskaya2016big,fang2015managing}.
\end{itemize}
Also,  
the corresponding algorithms have to be developed to overcome this heterogeneity in data acquisition. A meta-format of the data should be applied, so that all collected data can be represented within this format providing a common basis for the further analysis of data. 

The \emph{Data Quality Layer} is orthogonal to all the other Layers and deals with the quality of data that flows through these layers. Thus, it is dedicated to syntactical and semantic data analysis and verification, as well as quality assurance and usability aspects. 
One of important aspects is here also the tracking of data flows  through all layers  to provide means to reproduce data processing conducted in our architecture.

\begin{figure*}[ht!]
\begin{center}
\includegraphics[width=0.9\textwidth]{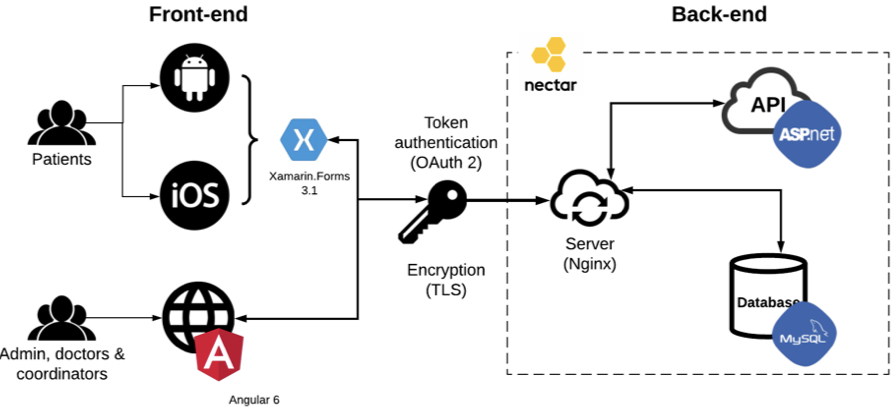}
\caption{CDP: Architecture of the hospital sub-domain}
\label{fig:arch2}
\end{center}
\end{figure*}

\vspace{5mm}

\begin{figure*}[hbt!]
\begin{center}
\includegraphics[width=\textwidth]{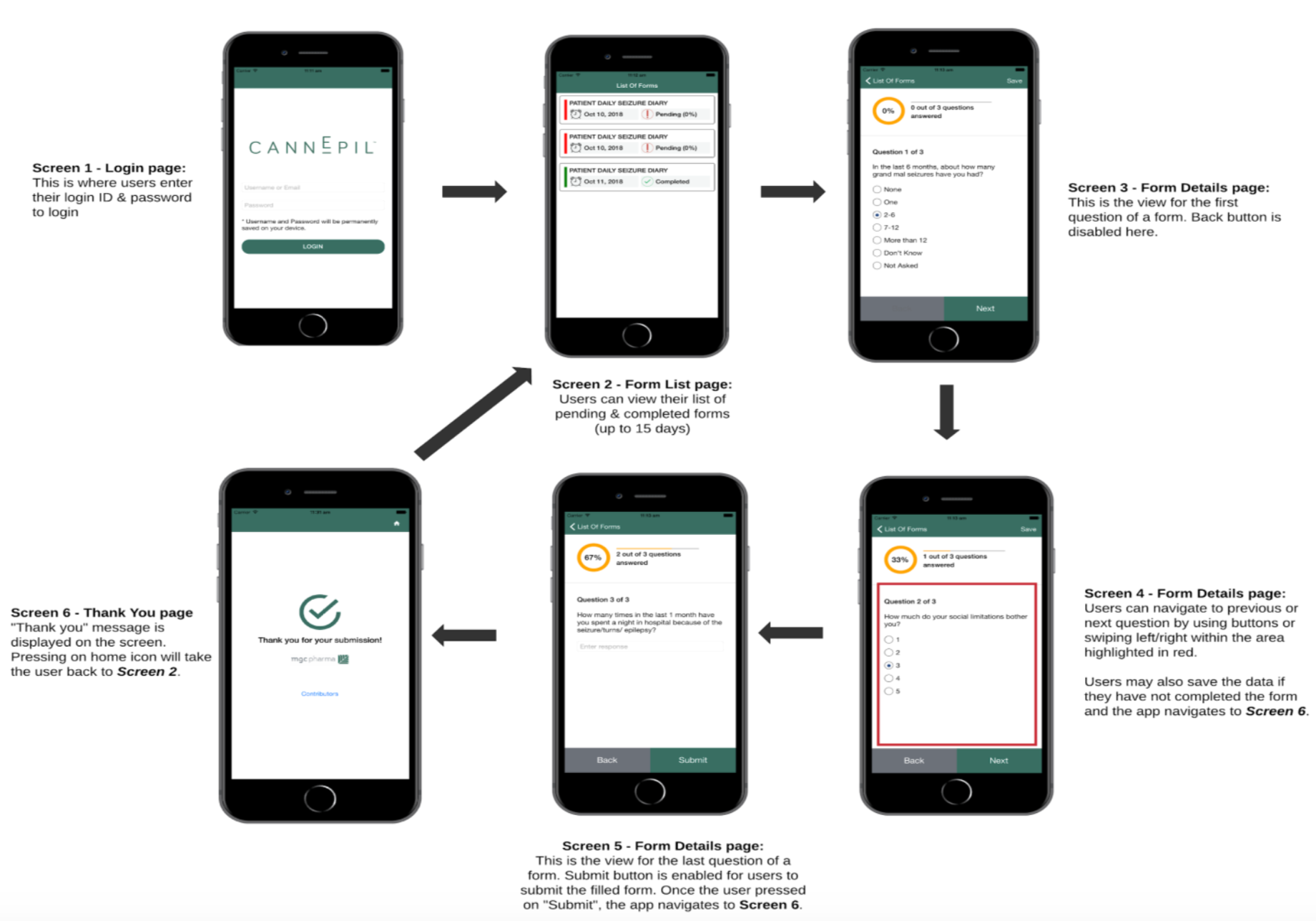} 
\caption{CDP: Flow Diagram for the hospital sub-domain (patient interface)}
\label{fig:mobile}
\end{center}
\end{figure*}

%===============================================
%\newpage  ~ \newpage ~ \newpage
\section{\uppercase{Cannabinoids Data Platform}}
\label{sec:impl}

In this section, we introduce the current implementation of CDP. 
Currently, the prototype focuses on the infrastructure required to collect data from the users within the hospital and research sub-domains: patients, doctors/ treatment coordinators and researchers. 
Figure \ref{fig:arch2} presents the architecture of the hospital sub-domain.

 \begin{figure*}[ht!]
\begin{center}
\includegraphics[width=0.7\textwidth]{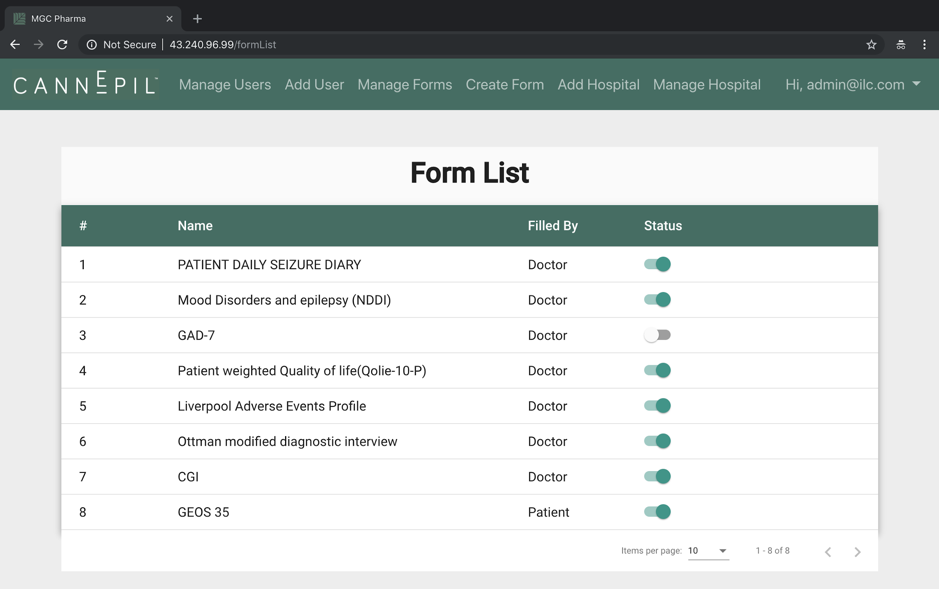}
\caption{CDP: User interface for the hospital sub-domain (``Manage Forms'' page, part of the functionality provided to doctor / treatment  coordinator users)}
\label{fig:doctor1}
\end{center}
\end{figure*}
 
\begin{figure*}[ht!]
\begin{center}
\includegraphics[width=0.7\textwidth]{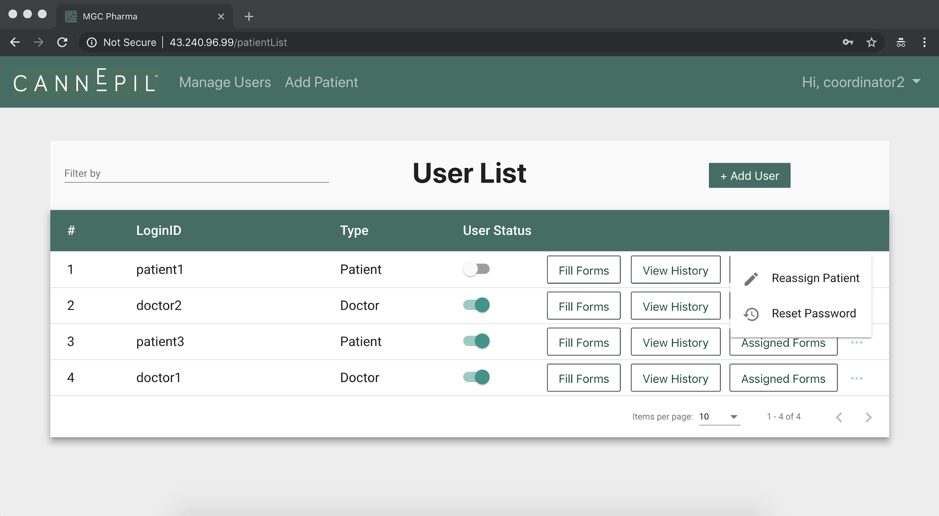}
\caption{CDP: User interface for the hospital sub-domain (``Manage Users'' page, part of the functionality provided to doctor/ treatment coordinator users)}
\label{fig:coord1}
\end{center}
\end{figure*}

The prototype has two core interface components:  
\begin{itemize}
    \item 
    mobile applications (both iOS and Android) developed for patients,  see Figure~\ref{fig:mobile}; the  applications were built using Xamarin\footnote{\url{https://visualstudio.microsoft.com/xamarin}}, which provides cross-platform compatibility between Android and iOS platforms; Having a C$\#$-shared codebase, developers can use Xamarin tools to write native Android, iOS, and Windows apps with native user interfaces and share code across multiple platforms.
    \item 
    Web applications providing interfaces for doctors/ coordinators and researchers was developed using Angular 6, which is  
     a TypeScript-based open-source front-end web application platform\footnote{\url{https://angular.io}} (see Figure~\ref{fig:doctor1}  
     for an example of the implemented Web interfaces).
\end{itemize}
Between the Front-end and Back-end, we implemented an authentication system using OAuth2 with refresh token grant type. In addition, the system is using Transport Layer Security (TLS) cryptographic protocols to provide communications security over a computer network.
In Back-end, MySQL Database is used to store all resource data, and we implemented an ASP.NET Web API project to communicate with DB. The whole back-end including API (Application Programming Interface) and data base are hosted in Nectar Cloud\footnote{\url{https://nectar.org.au/research-cloud}} that provides free cloud services for Australian Researchers.

Users have to register and log in before they are provided a suitable interface for them. Patients can add their treatment history including severity of condition and effectiveness of particular formulations. They can keep track of their treatment history to understand what works for their condition. Their assigned doctors have access to their treatment data to customize the treatment.

Doctors can manage all their patient's cases, allocate to them corresponding questionnaires / forms to fill out on regular basis to collect data on the progress of the treatment and its effectiveness. Doctors can add comments and annotations for an individual case, as well as add/ remove treatments for the patients. %  

Researchers, users that have to request higher privileges in the system because they can browse any patient case to help their research. They can also search for a comprehensive investigation of strain data, which also includes added advanced search functionalities on the system.

The standard functionality to deal with the user profiles is also covered within the current version of the prototype: all users can 
update their profile; patients and doctors can submit a request to become a researcher with the CDP, where the researcher role provides an access to anonymised data on treatments and experiments being conducted.

%===============================================
\section{\uppercase{Conclusions}} 
\label{sec:conclusions}
In this paper, we presented the core ideas of the Cannabinoids Data Platform, which goal is to integrate data from the complete  cannabinoids domain, including research, hospital and growers sub-domains. 
Dealing with multiple heterogeneous sub-domains means additional challenges in  collecting and integrating heterogeneous and unstructured or semi-structured data from several sources, as we cannot rely on a single data structure/ format or predefined data standards. However, a meta-structure/format can be introduces to integrate the data.  
We implemented a prototype of the platform, which currently has  two types of interfaces:
 iOS and Android mobile applications developed for patients,  and
 Web applications developed for doctors/ treatment coordinators and researchers. The user interfaces in different sub-domains also differ, as we have to take int account not only the type of data we collect from each sub-domain but also the preferences and skills of the users within the sub-domain.

\newpage ~  \newpage ~  \newpage 
\bibliographystyle{apalike}

\vfill
\end{document}